\documentclass[secnumarabic, graphics,floatfix, nofootinbib,tightenlines,nobibnotes, aps, prl, 12pt]{revtex4-1}
\usepackage{graphicx}
\usepackage[english]{babel}
\usepackage[utf8]{inputenc}
\usepackage{mathrsfs}
\usepackage{scalerel,amsmath,amssymb}
\usepackage{amsfonts}
\usepackage{multirow}
\usepackage[section]{placeins}
\usepackage[center]{subfigure}

\begin{document}

 \newcommand{\bq}{\begin{equation}}
 \newcommand{\eq}{\end{equation}}
 \newcommand{\bqn}{\begin{eqnarray}}
 \newcommand{\eqn}{\end{eqnarray}}
 \newcommand{\nb}{\nonumber}
 \newcommand{\lb}{\label}

\title{Neutral regular black hole solution in generalized Rastall gravity}

\author{Kai Lin$^{1,2}$}\email{lk314159@hotmail.com}
\author{Wei-Liang Qian$^{2,3,4}$}\email{wlqian@usp.br}

\affiliation{$^{1}$ Hubei Subsurface Multi-scale Imaging Key Laboratory, Institute of Geophysics and Geomatics, China University of Geosciences, 430074, Wuhan, Hubei, China}
\affiliation{$^{2}$ Escola de Engenharia de Lorena, Universidade de S\~ao Paulo, 12602-810, Lorena, SP, Brazil}
\affiliation{$^{3}$ Faculdade de Engenharia de Guaratinguet\'a, Universidade Estadual Paulista, 12516-410, Guaratinguet\'a, SP, Brazil}
\affiliation{$^{4}$ Center for Gravitation and Cosmology, College of Physical Science and Technology, Yangzhou University, Yangzhou 225009, China}

\date{April. 29, 2019}

\begin{abstract}

In this work, we investigate the static, spherically symmetric regular black hole solutions in a generalized Rastall gravity.
In particular, the prescription of Rastall gravity implies that the present approach does not necessarily involve nonlinear electrodynamics.
Subsequently, the resulting regular black hole solution can be electrically and magnetically neutral.
The general properties of the regular black hole solutions are explored.
Moreover, specific solutions are derived and discussed, particularly regarding the parameter related to the degree of violation of energy-momentum conservation in Rastall theory.

\vspace{50pt}

\end{abstract}

\maketitle
\section{Introduction}
\renewcommand{\theequation}{1.\arabic{equation}} \setcounter{equation}{0}

The black hole is an interesting prediction of general relativity, a compact celestial body with enormous mass and immense gravity.
Since 2015, LIGO and Virgo collaborations have detected various gravitational waves emanated from the coalescence of black holes~\cite{agr-LIGO-01,agr-LIGO-02,agr-LIGO-03,agr-LIGO-04,agr-LIGO-06,agr-LIGO-07}, which serves as one of the most crucial evidence for the existence of black holes.
According to Penrose-Hawking singularity theorems~\cite{book-wald-general-relativity}, singularity might be found accompanying black hole solutions.
In the vicinity of a singularity, as the spacetime curvature approaches and exceeds the Planckian values, the notion of classical spacetime ceases to be valid.
Fortunately, regarding the cosmic censorship hypothesis proposed by Penrose, it is understood that the singularity is typically hidden behind the event horizon.
The latter is a one-way membrane which prevents the singularity from affecting any observer located outside of the horizon. 
Nonetheless, the concept of singularity in black hole physics plays a vital role concerning the information problem~\cite{agr-bh-information-03,agr-bh-information-04}, as well as the final fate of a black hole, associated with the evaporation due to the Hawking radiation~\cite{agr-hawking-radiation-01,agr-hawking-radiation-02}.
Alternatively, a somewhat conservative approach is to pursue a regular and effective geometric description for the region of the black hole spacetime, which otherwise would be singular.
One might aim for a moderate framework for analyzing many crucial physical problems without introducing significant deviations from the standard model.
In this regard, in 1968, Bardeen proposed a regular black holes metric, where a region of nonsingular spacetime substitutes the central singularity~\cite{agr-metric-regular-bh-01}.
Meanwhile, the metric asymptotically coincides with that of the Schwarzschild solution at infinity.
Subsequently, following this line of thought, various models for regular black holes were proposed~\cite{agr-metric-regular-bh-04,agr-metric-regular-bh-05,agr-metric-regular-bh-06,agr-metric-regular-bh-07,agr-metric-regular-bh-08}.
Later, it was argued that the black hole solutions could be physically interpreted in terms of nonlinear electrodynamics with magnetic monopole~\cite{agr-metric-regular-bh-02,agr-metric-regular-bh-03}.
Further investigations concerning the quasinormal modes of regular black hole metrics have also been carried out~\cite{agr-qnm-regular-bh-01,agr-qnm-regular-bh-02,agr-qnm-asympototic-iteration-04,agr-qnm-regular-bh-04,agr-qnm-regular-bh-05}, and the corresponding metrics are shown to be stable against various types of perturbations.
However, nonlinear electrodynamics encounters several potential difficulties.
Firstly, astronomical objects are, by and large, electrically neutral, or do not carry a substantial amount of charges.
Secondly, while the (linear) quantum electrodynamics is one of the most accurately validated theories in physics, there is still a lack of strong experimental support for nonlinear electrodynamics.
Lastly, the constructed regular black hole solution usually requires the concept of the magnetic monopole, which is a hypothetical elementary particle.
Moreover, owing to cosmic inflation, only an insignificant amount of magnetic monopoles might persist in the observable universe.

Regarding the above discussions, the present study involves an attempt to study regular black hole solutions in the framework of Rastall theory.
The derived solutions can be electrically and magnetically neutral.
Rastall's gravity was proposed in 1972 by Rastall~\cite{agr-rastall-01}, as a generalization of Einstein's general relativity.
It is proposed that the conservation of energy-momentum tensor in curved spacetime can be relaxed, and it attains the form
 \bqn
 \lb{1.1}
T^\nu_{\mu;\nu}=a_\mu \ ,
 \eqn
where $a_\mu$ should vanish in flat spacetimes so that in this case the theory restores the Einstein's gravity.

In this original work, Rastall assumes 
\bqn
a_\mu=\lambda \nabla_\mu R \ , \lb{eqaa}
\eqn 
and therefore the field equation becomes
 \bqn
 \lb{1.2}
R_{\mu\nu}-\frac{1}{2}g_{\mu\nu}R=\kappa (T_{\mu\nu}-\lambda g_{\mu\nu}R) \ ,
 \eqn
where $\kappa=8\pi G/c^4$ and $\lambda$ is a constant.
As a theory of modified gravity, Rastall's gravity has received increasing attention lately~\cite{agr-rastall-02,agr-rastall-03,agr-rastall-04,agr-rastall-05,agr-rastall-06,agr-rastall-07,agr-rastall-09,agr-rastall-10,agr-rastall-11,agr-rastall-12,agr-rastall-13,agr-rastall-14,agr-rastall-15,agr-rastall-16,agr-rastall-17,agr-rastall-18,agr-rastall-19,agr-rastall-20,agr-rastall-21}, particuarly due to recent findings in cosmology~\cite{agr-rastall-cosmo-01,agr-rastall-cosmo-02,agr-rastall-cosmo-03,agr-rastall-cosmo-04,agr-rastall-cosmo-05,agr-rastall-cosmo-06,agr-rastall-cosmo-07,agr-rastall-cosmo-08,agr-rastall-cosmo-09,agr-rastall-cosmo-10}.

In a recent study~\cite{agr-rastall-21}, it is pointed out that in accordance to Rastall's original proposal, $a_\mu$ can adopt various forms besides Eq.(\ref{eqaa}).
This is because the only requirement is that $a_\mu$ vanishes in flat spacetimes~\cite{agr-rastall-01}, as one does not lead to any confliction with the present observations.
In this context, we assume 
\bqn
a_\mu=\nabla^\nu{\cal A}_{\nu\mu} \ ,
\eqn 
where ${\cal A}_{\nu\mu}={\cal A}_{\mu\nu}$, and the field equation can be written as
 \bqn
 \lb{1.3}
R_{\mu\nu}-\frac{1}{2}g_{\mu\nu}R=\kappa (T_{\mu\nu}-{\cal A}_{\mu\nu}) \ .
 \eqn
Here ${\cal A}_{\mu\nu}$ as well as its derivatives must be sufficiently small where the curvature of the spacetime vanishes.
It is straightforward to show that one can formally express various modified theories of gravity, such as $f(R)$ gravity and quadratic gravity, regarding the above-generalized form of Rastall's gravity~\cite{agr-rastall-18,agr-rastall-21}.

In this work, we investigate Rastall's gravity with
 \bqn
 \lb{1.4}
{\cal A}_{\mu\nu}=\lambda g_{\mu\nu}H(R) \ ,
 \eqn
where $H=H(R)$ is an arbitrary funcion of the Ricci scalar.
According to the above discussions, one requires that $H=0$ in flat spacetime, where $R=0$.
In the remainder of my paper, it is shown that one may derive regular black hole solutions for the Rastall's gravity determined by Eq.(\ref{1.4}).

The present paper is organized as follows.
In the next section, we discuss the general properties of regular black holes.
The specific form of the line element is given explicitly.
In section III, we provide a detailed account for constructing the regular black hole solution in Rastall's gravity which assumes Eq.(\ref{1.4}).
Further discussions and concluding remarks are given in the last section.

\section{Properties of regular black holes}

\renewcommand{\theequation}{2.\arabic{equation}} \setcounter{equation}{0}

In the section, we discuss the general properties of the metric for a regular black hole and derive the relevant requirements to be fulfilled.
We start by considering the following form of static, spherically symmetric metric in four-dimensional spacetimes,
 \bqn
 \lb{2.1}
ds^2=-f(r)dt^2+\frac{dr^2}{f(r)}+r^2(d\theta^2+\sin^2\theta d\varphi^2) \ ,
 \eqn
where
 \bqn
 \lb{2.2}
f(r)=1-\frac{2{\mathcal M}(r)}{r}=1-\frac{2M_0}{r}C_M(r) \ .
 \eqn
Here $M_0$ is considered as the mass of black hole as measured by an inertial observer sitting at infinity, and the parameter $C_M$ is introduced in the last equality which satisfies $C_M\rightarrow1$ as $r\rightarrow\infty$.
In what follows, we derive the condition for the black hole solution to be regular at $r=0$ in terms of that of $C_M(r)$.

The avoidance of singularity implies that the curvature of the spacetime does not diverge.
In practice, it is achived by investigating the convergence of the relavant scalar quantites, namely, $R=g^{\mu\nu}R_{\mu\nu}$, $\mathbb{R}=R_{\alpha\beta}R^{\alpha\beta}$ and $\mathscr{R}=R_{\alpha\beta\gamma\sigma}R^{\alpha\beta\gamma\sigma}$.
For a static regular black hole, we substitute Eq.(\ref{2.1}) and Eq.(\ref{2.2}) into the above quantities and find
 \bqn
 \lb{2.3}
 R&=&\frac{2M_0}{r^2}\left(2C_M'+rC_M''\right) \ ,\nb\\
 \mathbb{R}&=&\frac{2M_0^2}{r^4}\left(4C_M'^2+r^2C_M''^2\right) \ ,\nb\\
 \mathscr{R}&=&\frac{4M_0^2}{r^6}\left[12C_M^2+4rC_M\left(rC_M''-4C_M'\right)+r^2\left(r^2C_M''^2-4rC_M'C_M''+8C_M'^2\right)\right] \ ,
 \eqn
where $C_M'\equiv\frac{dC_M}{dr}$ and $C''_M\equiv \frac{d^2C_M}{dr^2}$. 
In the vicinity of $r=0$, we assume $C_M\sim r^{\alpha_\text{center}}+{\cal O}(r^{\alpha_\text{center}})$.
By substituting this condition into the above scalars and expanding around $r=0$, one finds that the requirement of convergence implies a condition ${\alpha_\text{center}}\ge3$.
On the other hand, one has $C_M\rightarrow 1$ as $r\rightarrow\infty$, which usually does not lead to any curvature divergence.
The above condition should be satisfied for any regular black hole solution.

Now, let us discuss the properties of the metric near the black hole horizon.
If one denotes the horizon by $r_p$, for $f(r_p)=0$ one has
 \bqn
 \lb{2.4}
f(r)=1-\frac{r_pC_M(r)}{rC_M(r_p)} \ .
 \eqn
Therefore, the condition $r_p > 0$ implies $C_M(r_p)>0$.
The temperature at the horizon is
 \bqn
 \lb{2.5}
T_H=\frac{1-r_p C_p}{4\pi r_p} \ .
 \eqn
where $C_p=\left.C_M'/C_M\right|_{r=r_p}$.

In literature, in order to derive a metric for the regular black hole, one introduces nonlinear electrodynamic field to the system.
Alternatively, in the following section, it is shown that a static regular black hole solution can be found in the Rastall's gravity in terms of $H(R)$.

\section{Regular black holes in Rastall's gravity}

\renewcommand{\theequation}{3.\arabic{equation}} \setcounter{equation}{0}

In static, spherically symmetric black hole metric, the energy-momentum tensor of a given type of fluid surrounding the black hole can be written as~\cite{agr-metric-quintessence-01,agr-rastall-02}
 \bqn
 \lb{3.1}
T^t_t&=&-\rho(r) \ ,\nb\\
T^i_j&=&-\rho(r)\alpha\left[\beta\delta^i_j-(1+3\beta)\frac{r_ir^j}{r_n r^n}\right] \ ,
 \eqn
where $\rho$ and $p$ are the energy density and pressure of the matter field.
When averaged over the angles for an isotropic system, the spatial components read 
 \bqn
 \lb{3.2}
\langle T^i_j\rangle=\frac{\alpha}{3}\rho\delta^i_j=p\delta^i_j \ ,
 \eqn
where we have used $\langle r^ir_j\rangle=\frac{1}{3}\delta^i_j r^n r_n$.
When considering barotropic equation of state, we have~\cite{agr-metric-quintessence-01,agr-rastall-02}
 \bqn
 \lb{3.3}
p&=&\omega \rho \ ,\nb\\
\omega&=&\frac{\alpha}{3} \ ,\nb\\
\beta&=&-\frac{1+3\omega}{6\omega}\ .
 \eqn
The corresponding energy-momentum tensor takes the following form
 \bqn
 \lb{3.4}
T^t_t&=&=T^r_r=-\rho(r) \ ,\nb\\
T^\theta_\theta&=&T^\varphi_\varphi=\frac{1}{2}(1+3\omega)\rho(r) \ .
 \eqn

Now, by substituting the static black hole metric, Eq.(\ref{2.1}), into the field equation of Rastall's theory, Eq.(\ref{1.3}) and Eq.(\ref{1.4}), one obtains
 \bqn
 \lb{3.5}
rf'(r)+f(r)-1+\kappa \left[r^2\rho(r)+\lambda r^2H\right]&=&0 \ ,\nb\\
rf''(r)+2f'(r)+\kappa\left[2\lambda r H-(1+3\omega) r \rho(r)\right]&=&0 \ ,
 \eqn
while the equation for the energy-momentum tensor, Eq.(\ref{1.1}), becomes
 \bqn
 \lb{3.6}
\rho'(r)+3\frac{1+\omega}{r}\rho(r)+\lambda \frac{d H}{d r}=0 \ .
 \eqn
We note Eq.(\ref{3.6}) is not an independent equation since it is implied by Eq.(\ref{3.5}).
Here, $H=H(R)$ is a function of the Ricci scalar, which determines the specific form of the metric while satisfying Eq.(\ref{3.6}).

By solving Eq.(\ref{3.5}-\ref{3.6}), one finds
 \bqn
 \lb{3.7}
H&=&-\frac{(1+3\omega)(f(r)-1)+3(1+\omega)rf'(r)+r^2f''(r)}{3r^2\lambda\kappa(1+\omega)} \ ,\nb\\
\rho(r)&=&\frac{r^2f''(r)-2f(r)+2}{3\kappa(1+\omega)r^2} \ .
 \eqn
It can be rewritten in terms of $C_M$ by making use of Eq.(\ref{2.2})
 \bqn
 \lb{3.8}
H&=&2M_0\frac{(1+3\omega)C_M'(r)+rC_M''(r)}{3\kappa\lambda(1+\omega)r^2} \ ,\nb\\
\rho(r)&=&2M_0\frac{2C_M'(r)-rC_M''(r)}{3\kappa(1+\omega)r^2} \ .
 \eqn

Now are in the position to study the conditions under which the above solution is indeed regular.
According to the discussions above, in flat spacetime, $H(R)$ satisfies the condition
 \bqn
 \lb{3.9}
H \rightarrow 0 \ \ \mathrm{as}\ \ R\rightarrow 0 \ .
 \eqn
This also implies that $H(R)\rightarrow 0$ as $r\rightarrow \infty$.
On the other hand, it is also required that $\rho(r)$ does not possess any singularity in the entire range $r\in[0,+\infty)$.
By making use of the properties of the curvature scalars and $C_M(r)$ discussed previously, we find the desired conditions
 \bqn
 \lb{3.10}
C_M(r\rightarrow 0)\rightarrow r^{\alpha_{\text{center}}}~~&\text{with}&~~\alpha_{\text{center}}\ge 3 \ ,\nb\\
C_M(r\rightarrow \infty)\rightarrow 1+r^{\alpha_{\text{infinity}}}~~&\text{with}&~~\alpha_{\text{infinity}}<0 \ .
 \eqn

The above conditions, Eq.(\ref{3.10}), can be satisfied when $C_M$ possesses the form of a fraction, where both numerator and denominator are polynomials in $r$.
The first line of Eq.(\ref{3.10}) dictates that the lowest degree of the monomials on the numerator is at least three orders larger than that on the denominator.
The second line of Eq.(\ref{3.10}) implies that the highest degree of the monomials on the numerator must be smaller than that on the denominator.
As a simple illustration, a possible solution reads
 \bqn
 \lb{3.11}
C_M(r)=\frac{r^3}{r^3+2\sigma^2} \ ,
 \eqn
where $\sigma$ is a constant.
If $\sigma= 0$, the corresponding metric is not regular.
For $\sigma\ne 0$, we have a regular black hole solution
 \bqn
 \lb{3.12}
f(r)&=&1-\frac{2M_0r^2}{r^3+2\sigma^2} \ ,\nb\\
\rho(r)&=&\frac{24M_0\sigma^2r^3}{\kappa(1+\omega)(r^3+2\sigma^2)^3} \ ,\nb\\
H&=&12M_0\sigma^2\frac{(\omega-1)r^3+2(\omega+1)\sigma^2}{\kappa\lambda(1+\omega)(r^3+2\sigma^2)^3}\ ,\nb\\
R&=&24M_0\sigma^2\frac{4\sigma^2-r^3}{(r^3+2\sigma^2)^3} \ .
 \eqn
In fact, the above solution can readily be identified as a Hayward regular black hole~\cite{agr-metric-regular-bh-04}.
Moreover, the above results imply that $r=r(R)$.
Subsequently, one finds the following relations
 \bqn
 \lb{3.13}
H(R)=3{\cal B}^{2}R^2\frac{(\omega-1)\left[{\cal B}^2M_0^{1/3}-2RM_0^{2/3}\right]-2\sigma{\cal B}R}{\kappa\lambda(1+\omega)({\cal B}^{2}-2RM_0^{1/3})^3} \ ,
 \eqn
where ${\cal B}=\left(R^{3/2}\sqrt{81\sigma^2R+8M_0}-9\sigma R^2\right)^{1/3}$.
Obviously, Eq.(\ref{3.13}) satisfies the condition that $H(R)\to 0$ as $R\rightarrow 0$.
Therefore, we have constructed regular black hole solutions in Rastall's gravity, and in particular, we note that $\omega\not=-1$.

\section{Further discussions and concluding remarks}

\renewcommand{\theequation}{4.\arabic{equation}} \setcounter{equation}{0}

Following the original ideal proposed by Rastall, in this work, the conservation of the energy-momentum tensor is generalized to the form
 \bqn
 \lb{4.1}
\nabla_\mu T^\mu_\nu=\lambda \nabla_{\nu} H(R) \ ,
 \eqn
while the corresponding field equation reads
 \bqn
 \lb{4.2}
R_{\mu\nu}-\frac{1}{2}g_{\mu\nu}R=\kappa\left(T_{\mu\nu}-\lambda g_{\mu\nu}H(R)\right) \ ,
 \eqn
where $H(R)$ vanishes in flat spacetime.

In previous sections, we have shown that regular black hole solutions can be derived in the framework of the Rastall's gravity.
As discussed above, the obtained black hole spacetimes can be electrically neutral, which does not involve nonlinear electrodynamics as well as the associated theoretical speculations.
In this context, the present study provides a novel possibility for models of regular black holes.

The constructed black hole spacetime is surrounded by a matter field, described by energy-momentum tensor, Eq.(\ref{3.4}).
However, to guarantee that the solution is nonsingular, one finds that the equation of state of the matter field has to satisfy the condition $\omega\not=-1$.
In other words, the dark energy model in terms of cosmological constant cannot be a candidate for hosting a regular black hole solution in question.

On the other hand, when assuming $T_{\mu\nu}=0$, the static black hole solution found in the present model is no longer regular.
However, interestingly enough, in this case one can show that the resulting metric is equivalent to an (anti-)de Sitter spacetime.
To demonstrate this point, one first contracts both sides of Eq.(\ref{4.2}) by $g^{\mu\nu}$ to obtain
 \bqn
 \lb{4.3}
4\lambda\kappa H(R)-R=0 \ .
 \eqn
This is an algebraic equation, and in general, it possesses a nonvanishing root of $R$, besides the one at the origin.
Thus one may rewrite Eq.(\ref{4.2}) as follows
 \bqn
 \lb{4.4}
R_{\mu\nu}=g_{\mu\nu}\left(\frac{1}{2}R-\kappa\lambda H(R)\right)\equiv g_{\mu\nu}\Lambda_{\text{eff}} \ ,
 \eqn
where $R$ is a non-zero root of Eq.(\ref{4.3}).
For the reason which will shortly become clear, $\Lambda_{\text{eff}}$ is defined as the effective cosmolocial constant, and one has
 \bqn
 \lb{4.5}
R&=&4\Lambda_{\text{eff}}\ ,\nb\\
H&=&\frac{\Lambda_{\text{eff}}}{\kappa\lambda} \ .
 \eqn
As a result, the metric Eq.(\ref{2.1}) can be written as
 \bqn
 \lb{4.6}
f(r)=1-\frac{2M}{r}-\frac{\Lambda_{\text{eff}}}{3}r^2 \ .
 \eqn
Even though the cosmological constant is not an {\it {ad hoc}} assumption in the field equation, the properties of the resulting metric indicate that the space is de Sitter.
Therefore it arises naturally from the consistency of the theory.
For instance, let us consider the case where $\Lambda_{\text{eff}} >0$.
The de Sitter background is realized in the context of the present generalized Rastall gravity in terms of the effective cosmological constant $\Lambda_{\text{eff}}=\frac{R}{4}$.
If one chooses $H(R)=R^n$ with $n>1$ and $\lambda>0$, we find
 \bqn
 \lb{4.7}
R&=&4\Lambda_{\text{eff}}=\left(\frac{1}{4\kappa\lambda}\right)^{\frac{1}{n-1}} \ ,\nb\\
H&=&\frac{\Lambda_{\text{eff}}}{\kappa\lambda}=\left(\frac{1}{4\kappa\lambda}\right)^{\frac{n}{n-1}}\ ,\nb\\
\Lambda_{\text{eff}}&=&\frac{1}{4}\left(\frac{1}{4\kappa\lambda}\right)^{\frac{1}{n-1}} \ .
 \eqn

Moreover, it is not difficult to show that one can further extend the above considerations to rotating black holes with the presence of the (linear) Maxwell field.
In this cae, gravitational field equation is given by
 \bqn
 \lb{4.A}
R_{\mu\nu}-\frac{1}{2}g_{\mu\nu}R+\kappa \lambda g_{\mu\nu}R-2F_{\mu\alpha}F^\alpha_\nu+\frac{1}{2}g_{\mu\nu}F_{\alpha\beta}F^{\alpha\beta}=0 ,
 \eqn
where the electromagnetic tensor $F_{\mu\nu}=\partial_\mu A_\nu-\partial_\nu A_\mu$ satisfies the Maxwell equations
 \bqn
 \lb{4.B}
\partial_\mu\left(\sqrt{-g}F^{\mu\nu}\right)=0 .
 \eqn
We note that Eq.(\ref{4.A}) still falls back to Eq.(\ref{4.3}) when contracting both sides by $g^{\mu\nu}$, as the two terms involving electromagentic tensor cancel out. 
This implies that we again arrive at an equation similar to Eq.(\ref{4.4})
 \bqn
 \lb{4.C}
R_{\mu\nu}-2F_{\mu\alpha}F^\alpha_\nu+\frac{1}{2}g_{\mu\nu}F_{\alpha\beta}F^{\alpha\beta}=g_{\mu\nu}\left(\frac{1}{2}R-\kappa\lambda H(R)\right)\equiv g_{\mu\nu}\Lambda_{\text{eff}} \ ,
 \eqn
where $R$ and $H(R)$ still satisfy Eq.(\ref{4.5}). 

This result furnishes a Kerr-Newman (Anti-)de Sitter black hole metric in Rastall gravity, namely,
 \bqn
 \lb{4.D}
ds^2=-\frac{\Delta_r}{\Xi^2\rho^2}\left(dt-a\sin^2\theta d\varphi\right)^2+\frac{\Delta_\theta}{\Xi^2\rho^2}\sin^2\theta\left(adt-\left(r^2+a^2\right)d\varphi\right)^2+\frac{\rho^2}{\Delta_r}dr^2+\frac{\rho^2}{\Delta_\theta}d\theta^2,\nb\\
 \eqn
where the electromagnetic potentials $A_\mu=\frac{Qr}{\Xi\rho^2}\left(\delta^t_\mu-a\sin^2\theta\delta^\varphi_\mu\right)$, and $Q$ and $a$ are electrical charge and angular momentum per unit mass, respectively.
Also, 
 \bqn
 \lb{4.E}
\rho^2&=&r^2+a^2\cos^2\theta,\nb\\
\Xi&=&1+\frac{a^2}{3}\Lambda_{\text{eff}} ,\nb\\
\Delta_r&=&\left(1-\frac{r^2}{3}\Lambda_{\text{eff}}\right)\left(r^2+a^2\right)-2Mr+Q^2 ,\nb\\
\Delta_\theta&=&1+\frac{a^2\cos^2\theta}{3}\Lambda_{\text{eff}}  .
 \eqn

The above (Anti-)de Sitter black hole solutions are physically intriguing, since it emerges entirely from the vacuum Rastall equation without the cosmological constant, which probaly may lead to further implications in cosmology.
More effort alone this line of thought is under progress.

To summarize, we studied the static, spherically symmetric black hole solutions in a generalized Rastall gravity.
An essential feature of the derived solutions is that the black holes can be electrically and magnetically neutral, which is distinct from most literature on this topic where the nonlinear electrodynamic field is involved. 
We discussed the general properties of the regular black hole solutions.

\section*{\bf Acknowledgements}

We gratefully acknowledge the financial support from National Natural Science Foundation of China (NNSFC) under contract No.11805166,
as well as Brazilian funding agencies
Funda\c{c}\~ao de Amparo \`a Pesquisa do Estado de S\~ao Paulo (FAPESP),
Conselho Nacional de Desenvolvimento Cient\'{\i}fico e Tecnol\'ogico (CNPq),
and Coordena\c{c}\~ao de Aperfei\c{c}oamento de Pessoal de N\'ivel Superior (CAPES).

\bibliographystyle{h-physrev}
\bibliography{references_qian}

\end{document}